# United Nations Basic Space Science Initiative (UNBSSI) 1991-2012 and Beyond

**A.M. Mathai**
**Centre for Mathematical and Statistical Sciences, Peechi Campus, KFRI, Peechi-680653, Kerala, India, and Department of Mathematics and Statistics, McGill University, Montreal, Canada, H3A 2K6, mathai@math.mcgill.ca**

**H.J. Haubold**
**Centre for Mathematical and Statistical Sciences, Peechi Campus, KFRI, Peechi-680653, Kerala, India, and Office for Outer Space Affairs, United Nations, Vienna International Centre, P.O. Box 500, 1400 Vienna, Austria, hans.haubold@gmail.com**

**W.R. Balogh**
**Office for Outer Space Affairs, United Nations, Vienna International Centre, P.O. Box 500, 1400 Vienna, Austria, werner.balogh@unoosa.org**

## Abstract

The present paper contains an overview and summary on the achievements of the basic space science initiative in terms of donated and provided planetariums, astronomical instruments, and space weather instruments, particularly operating in developing nations. These instruments have been made available to respective host countries, particularly developing nations, through the series of twenty basic space science workshops, organized through the United Nations Programme on Space Applications since 1991.

Organized by the United Nations, the European Space Agency (ESA), the National Aeronautics and Space Administration (NASA) of the United States of America, and the Japan Aerospace Exploration Agency (JAXA), the basic space science workshops were organized as a series of workshops that focused on basic space science (1991-2004), the International Heliophysical Year 2007 (2005-2009), and the International Space Weather Initiative (2010-2012) proposed by the Committee on the Peaceful Uses of Outer Space on the basis of discussions of its Scientific and Technical Subcommittee, as reflected in the reports of the Subcommittee. Workshops on the International Space Weather Initiative in the series were hosted by the Government of Egypt in 2010 (see A/AC.105/994), the Government of Nigeria in 2011, and the Government of Ecuador in 2012 (see A/AC.105/1030). Workshops on the International Heliophysical Year 2007 were hosted by the United Arab Emirates in 2005 (see A/AC.105/856), India in 2006 (see A/AC.105/882), Japan in 2007 (see A/AC.105/902), Bulgaria in 2008 (see A/AC.105/919) and the Republic of Korea in 2009 (see A/AC.105/964). Workshops on basic space science were hosted by the

Governments of India (see A/AC.105/489), Costa Rica and Colombia (see A/AC.105/530), Nigeria (see A/AC.105/560/Add.1), Egypt (see A/AC.105/580), Sri Lanka (see A/AC.105/640), Germany (see A/AC.105/657), Honduras (see A/AC.105/682), Jordan (see A/AC.105/723), France (see A/AC.105/742), Mauritius (see A/AC.105/766), Argentina (see A/AC.105/784) and China (see A/AC.105/829).

All workshops were co-organized by the International Astronomical Union (IAU) and the Committee on Space Research (COSPAR).

## 1. Introduction

The basic space science initiative was a long-term effort for the development of astronomy and space science through regional and international cooperation in this field on a worldwide basis, particularly in developing nations. Basic space science workshops were co-sponsored and co-organized by ESA, JAXA, and NASA.

A series of workshops on basic space science was held from 1991 to 2004 (India 1991, Costa Rica and Colombia 1992, Nigeria 1993, Egypt 1994, Sri Lanka 1995, Germany 1996, Honduras 1997, Jordan 1999, France 2000, Mauritius 2001, Argentina 2002, and China 2004; http://neutrino.aquaphoenix.com/un-esa/) and addressed the status of astronomy in Asia and the Pacific, Latin America and the Caribbean, Africa, and Western Asia. Through the lead of Professor Dr. Masatoshi Kitamura (1926-2012) from the National Astronomical Observatory Japan, astronomical telescope facilities were inaugurated in seven developing nations and planetariums were established in twenty developing nations based on the donation of respective equipment by Japan.

Pursuant to resolutions of the Committee on the Peaceful Uses of Outer Space of the United Nations (COPUOS) and its Scientific and Technical Subcommittee, since 2005, these workshops focused on the preparations for and the follow-ups to the International Heliophysical Year 2007 (UAE 2005, India 2006, Japan 2007, Bulgaria 2008, South Korea 2009; www.unoosa.org/oosa/SAP/bss/ihy2007/index.html). IHY's legacy is the current operation of 16 worldwide instrument arrays with close to 1000 instruments recording data on solar-terrestrial interaction from coronal mass ejections to variations of the total electron content in the ionosphere (http://iswi-secretariat.org/). Instruments are provided to hosting institutions by entities of Armenia, Brazil, France, Israel, Japan, Switzerland, and the United States.

Starting in 2010, the workshops focused on the International Space Weather Initiative (ISWI) as mandated in a three-year-work plan as part of the deliberations of COPUOS. Workshops on ISWI were scheduled for Egypt in 2010 for Western Asia, Nigeria in 2011 for Africa, and Ecuador in 2012 for Latin America and the Caribbean.

Under the leadership of Professor K. Yumoto, the International Center for Space Weather Science and Education at Kyushu University, Fukuoka, Japan (www.serc.kyushu-u.ac.jp/index_e.html), was established through the basic space science initiative in 2012. The Centre for Basic Space Research, lead by Professor P.N. Okeke at the University of Nigeria in Nsukka, was established in Nigeria (www.unn.edu.ng/centres/centre-basic-space-research). The Centre for Mathematical and Statistical Sciences, lead by Professor A.M. Mathai and located in Kerala, India, supported all basic space science workshops (www.cmsintl.org).

Activities of basic space science initiative were also coordinated with the Regional Centres for Space Science and Technology Education, affiliated to the United Nations

([www.unoosa.org/oosa/en/SAP/centres/index.html](www.unoosa.org/oosa/en/SAP/centres/index.html)) and the International Committee on Global Navigation Satellite Systems (www.unoosa.org/oosa/en/SAP/gnss/icg.html).

**International Heliophysical Year 2007 (IHY 2007) and International Year of Astronomy 2009 (IYA 2009)**

In 2007, a number of major anniversaries occurred, among them the 50th anniversary of the International Geophysical Year, the launch of Sputnik 1, and the 50th session of the United Nations Committee on the Peaceful Uses of Outer Space (UNCOPUOS). Particularly IHY 2007 was an opportunity to (i) advance the understanding of the fundamental heliophysical processes that govern the Sun, Earth, and heliosphere, (ii) continue the tradition of international research and advancing the legacy of IHY 1957, and (iii) demonstrate the beauty, relevance and significance of space and Earth science to the world (http://ihy2007.org). Observing and implementing the mission and vision of IHY 2007, the basic space science initiative, in cooperation with ESA, NASA, JAXA, COSPAR, IAU, and the IHY Secretariat, hold international workshops in the United Arab Emirates in 2005 (www.ihy.uaeu.ac.ae/), in India in 2006 (www.iiap.res.in/ihy), in Japan in 2007 (http://solarwww.mtk.nao.ac.jp/UNBSS_Tokyo07/), and in Bulgaria in 2008 (http://newserver.stil.bas.bg/SUNGEO). The starting date of IHY 2007 was set to February 19, 2007. On that date, during the session of the Scientific and Technical Subcommittee of UNCOPUOS, the IHY kick-off included an IHY exhibit, press briefing, and an opening ceremony in the United Nations Office Vienna (http://ihy2007.org). The regional coordinators, Steering Committee members, and Advisory Committee members participated in the IHY kick-off event. The Austrian Academy of Sciences hosted a one-day symposium on IHY 2007 in Vienna on 20 February 2007.

The workshop for 2009 was hosted by the Republic of Korea (http://ihy.kasi.re.kr/meeting.php). This workshop also covered thematic areas as pursued by the International Year of Astronomy 2009 (www.astronomy2009.org). The basic space science initiative contributed to the Astronomy for the Developing World: Strategic Plan 2010-2020 ([http://iau.org/static/education/strategicplan_2010-2020.pdf](http://iau.org/static/education/strategicplan_2010-2020.pdf) ).

## 2. Astronomical telescopes donated to developing countries through the Official Development Assistance programme of Japan

In order to promote education and research in developing nations, the Government of Japan has provided developing nations with high-grade equipment under the framework of the Official Development Assistance (ODA) cooperation programme since 1982 (see Table 1.). Under this cooperation programme, astronomical instruments have been donated to seven developing nations. The instruments donated included university-level reflecting telescopes together with various accessories. Table 2 describes ODA donations with the assistance of the

Japan International Cooperation Agency (JICA) and the cooperation with the Programme on Space Applications of the Office for Outer Space Affairs of the United Nations (OOSA).

The number of science students in developing nations is rapidly increasing. In addition, selected students attend Ph.D. courses at universities or science institutes in order to further pursue higher education and research. Many of these scholars are aware of the fact that the present age is called the "space age" and become therefore interested in the subject of space science and technology.

Similarly, the number of highly educated professionals in astronomy is also steadily increasing in developing nations; however, many developing nations do not have the adequate astronomical equipment so urgently needed for education and research purposes that such professionals could use. One example of the need to support cooperation programmes providing adequate astronomical equipment to developing nations is that old-fashioned refracting telescopes are still used in a good number of nations. There is still a great lack of modern high-grade reflecting telescopes of higher quality and better resolution that could be used to better observe astronomical phenomena.

Table 1

| | Receiving Institution | Location | Model | Option | Country | Year |
|---|---|---|---|---|---|---|
| 1. | Science Centre | Singapore | 40-cm Reflector | .. | Singapore | 1987 |
| 2. | Bosscha Observatory Institute of Technology | Bandung, Lembang, 40391 Java, Indonesia | 45-cm Cassegrain | Photoelectric photometer, spectrograph | Indonesia | 1988 |
| 3. | Chulalongkorn University | Physics Department Faculty of Science Bangkok 10330, Thailand | 45-cm Cassegrain | Photoelectric photometer, spectrograph | Thailand | 1989 |
| 4. | Arthur C. Clark Center for Modern Technologies | Colombo, Katubedda Moratuwa, Sri Lanka | 45-cm Cassegrain | Photoelectric photometer, spectrograph | Sri Lanka | 1995 |
| 5. | Facultad Politecnica Asuncion University | Campus Universitario, Observatorio, Astronomico, San Lorenzo Asunción, Paraguay | 45-cm Cassegrain | Photoelectric photometer, charge-coupled device | Paraguay | 1999 |
| 6. | Philippine Atmoshperic, Geophysical and Astronomical Services Administration | 1424 ATB Bldg., Quezon Avenue, 1104 Quezon City, Philippines | 45-cm Cassegrain | Photoelectric photometer, spectrograph | Philippines | 2000 |
| 7. | Cerro Calan Astronomical Observatory Universidad de Chile Departamento de Astronomia | Casilla 36-D, Santiago, Chile | 45-cm Cassegrain | Charge-coupled device | Chile | 2001 |

## 3. Planetarium equipment donated to developing countries through the Official Development Assistance programme of Japan

Planetariums are important tools for education in astronomy. Nevertheless, only a limited number of developing nations have available planetariums located in their capital cities. On the other hand, industrialized nations have built a considerable number of planetariums that are used for space education not only in their capital cities, but also in villages, schools, and other places.

In order to promote education in developing nations, the Government of Japan has provided developing nations with equipment under the framework of the Official Development Assistance (ODA) cooperation programme since 1982 (see Table 2). Under cooperation programme, planetarium instruments have been donated to 20 developing nations. The instruments donated included complete planetariums used for educational purposes together with various accessories. Table 1 describes the ODA donations provided with the assistance of the Japan International Cooperation Agency (JICA) and the cooperation with the Programme on Space Applications of the Office for Outer Space Affairs of the United Nations (OOSA).

Table 2

| | Receiving Institution | Location | Model | Dome diameter (metres) | Seats | Country | Year |
|---|---|---|---|---|---|---|---|
| 1. | Pagoda Cultural Center | Yangon, Myanmar | GX | 12 | .. | Myanmar | 1986 |
| 2. | Haya Cultural Centre for Child Development | Post. B. 35022, Amman, Jordan | GEII-T | 6.5 | .. | Jordan | 1989 |
| 3. | National Planetarium Space Science Education Center | 53 Jalan Perdana, 50480 Kuala Lumpur, Malaysia | Minolta Infinium β | 20 | 213 | Malaysia | 1989 |
| 4. | Planetarium | Padre Burgos St., Ermita, Rizal Park, 2801 Manila, Philippines | GM-15s auxiliary projectors | 16 | 310 | Philippines | 1990 |
| 5. | Meghnad Saha Planetarium | University of Burdwan, Golapbag Burdwan-713104, West Bengal, India | GS-AT | 8.5 | 90 | India | 1993 |
| 6. | Planetario de la Ciudad de Buenos Aires "Galileo Galilei" | Av. Sarmiento y Belisario Roldán, s/n C1425FHA, Buenos Aires, Argentina | Auxiliary projectors | .. | 345 | Argentina | 1993 |
| 7. | Planetario de la Ciudad | Intendencia Municipal de Montevideo, Rivera 3245, 11600 Montevideo, Uruguay | Auxiliary projectors | .. | .. | Uruguay | 1994 |
| 8. | Ho-Chi Minh Memorial Culture Hall Vinh City Planetarium | Vinh University, No. 6 Le Mao Street, Vinh City, Nghee An Province, Viet Nam | GS | 8.5 | 80 | Viet Nam | 1998 |
| 9. | Planetarium | Science Center for Education, 928 Sukhumvit Road, Klong toey, Bangkok, 10110 Thailand | Auxiliary projectors | .. | .. | Thailand | 1998 |
| 10. | Planetarium | Ministry of Science and Technology, 255 Stanley Wijesundara, Mawatha, Colombo 7, Sri Lanka | Auxiliary projectors | .. | .. | Sri Lanka | 1998 |
| 11. | Tamilnadu Science and Technology Centre Anna Science Centre Planetarium | Pudukkottai National Highway, Near Tiruchirappalli Airport, Tiruchirappalli 620 007, India | GS | 8.5 | 90 | India | 1998 |
| 12. | Planetarium | City Park, ul. Chamzy 6, | .. | .. | .. | Uzbekistan | 2000 |

| | Receiving Institution | Location | Model | Dome diameter (metres) | Seats | Country | Year |
|---|---|---|---|---|---|---|---|
| | | Tashkent, Uzbekistan | | | | | |
| 13. | Planetario Padre Buenaventura Suárez S.J. | Oliva No. 479, Asunción, Paraguay | EX-3 | 5 | 23 | Paraguay | 2001 |
| 14. | Planetario Municipal | Florencia Astudillo y Alfonso Cordero, Parque de la Madre, Cuenca, Ecuador | .. | .. | 70 | Equador | 2002 |
| 15. | El Pequeño Sula, Museo para la Infancia of the City Hall of San Pedro Sula | Bulevar del Sur, Contiguo al Gimnasio Municipal, San Pedro Sula, Honduras C.A. | GS-T | 8.5 | .. | Honduras | 2003 |
| 16. | National Costa Rica University | San José, Costa Rica | GS-S | 8.5 | 40 | Costa Rica | 2003 |
| 17. | Laboratorio Central del Instituto Geofisico | Calle Badajoz 169-171, IV Etapa Mayorazgo, ATE, Lima 03, Perú | GS-T | 7.5 | .. | Perú | Scheduled for 2007 |
| 18. | National Astronomical Observatory of Tarija | Loc. Santa Ana Tarija, P.O. Box 346, Bolivia | GS-S | 8.5 | .. | Bolivia | Scheduled for 2008 |
| 19. | National History Museum | Havana, Cuba | .. | .. | .. | Cuba | Scheduled for 2007 |
| 20. | Tin Marín Children's Museum | Sexta Decima Calle Poniente, Centro Gimnacio Nacional y Parque Cuscatlan, San Salvador, El Salvador | GE-II | 6.5 | .. | El Salvador | 2007 |

## 4. Countries hosting instruments of the International Space Weather Initiative provided by Armenia, Brazil, France, Israel, Japan, Switzerland, and the United States of America

A major thrust of IHY 2007 and ISWI was to deploy arrays of small, inexpensive instruments such as magnetometers, radio telescopes, GPS receivers, all-sky cameras, etc. around the world to allow global measurements of ionospheric and heliospheric phenomena. The small instrument programme was envisioned as a partnership between instrument providers and instrument hosts in respective nations (see Table 3). The lead scientist is providing the instruments (or fabrication plans for instruments) in the array; the host nation provides human resources, facilities, and operational support to obtain data with the instrument, typically at a local university. Financial resources were not available through IHY and ISWI to build the instruments; these had to be obtained through the normal proposal channels. All instrument operational support for local scientists, facilities, data acquisition, etc. is provided by the host nation. The IHY and ISWI facilitated the deployment of several of these networks world-wide. Existing data bases and relevant software tools were identified to promote space science activities in developing nations. Extensive data on space science had been accumulated by a number of space missions. Similarly, long-term data bases are available from ground-based observations.

Table 3

01. **Algeria** (7) *AMBER(1), AWESOME(1), CHAIN(1), GPS_Africa(1), MAG_Africa(1), SID(2)*
02. **Antarctica** (2) *AWESOME(1), SID(1)*
03. **Argentina** (1) *SAVNET(1)*
04. **Armenia** (3) *SEVAN(3)*
05. **Australia** (18) *CALLISTO(2), GMDN(1), MAGDAS(11), OMTIs(1), SID(3)*
06. **Austria** (3) *CALLISTO(1), SID(2)*
07. **Azerbaijan** (3) *AWESOME(1), SID(2)*
08. **Bangladesh** (1) *SID(1)*
09. **Belgium** (1) *CALLISTO(1)*
10. **Benin** (1) *GPS_Africa(1)*
11. **Bosnia-Herzegovina** (1) *SID(1)*
12. **Botswana** (1) *GPS_Africa(1)*
13. **Brazil** (24) *CALLISTO(2), CSSTE(1), GMDN(1), MAGDAS(2), RENOIR(2), SAVNET(6), SCINDA(3), SID(7)*
14. **British Virgin Islands** (1) *SID(1)*
15. **Bulgaria** (2) *SEVAN(1), SID(1)*
16. **Burkina Faso** (3) *GPS_Africa(2), SID(1)*
17. **CHINA** (3) *SID(3)*
18. **Cameroon** (2) *AMBER(1), SCINDA(1)*
19. **Canada** (20) *MAGDAS(1), OMTIs(2), SID(17)*
20. **Cape Verde** (2) *GPS_Africa(1), SCINDA(1)*
21. **Central African Republic** (1) *MAG_Africa(1)*
22. **Chile** (1) *SCINDA(1)*
23. **China** (26) *SID(26)*
24. **Colombia** (8) *SCINDA(1), SID(7)*
25. **Congo** (7) *SCINDA(3), SID(4)*
26. **Costa Rica** (1) *CALLISTO(1)*
27. **Cote d'Ivoire** (4) *MAGDAS(1), MAG_Africa(2), SCINDA(1)*
28. **Croatia** (6) *SEVAN(1), SID(5)*
29. **Cypress** (1) *SID(1)*
30. **Czech Republic** (2) *CALLISTO(1), SID(1)*
31. **D Rep of Congo** (2) *SID(2)*
32. **Denmark** (2) *SID(2)*
33. **Djibouti** (1) *SCINDA(1)*
34. **Ecuador** (1) *MAGDAS(1)*
35. **Egypt** (9) *AWESOME(1), CALLISTO(1), CIDR(1), MAGDAS(2), SCINDA(1), SID(3)*
36. **England** (1) *SID(1)*
37. **Ethiopia** (20) *AMBER(1), AWESOME(1), MAGDAS(1), MAG_Africa(1), SCINDA(2), SID(14)*
38. **Fiji** (1) *AWESOME(1)*
39. **Finland** (2) *CALLISTO(2)*
40. **France** (4) *SID(4)*
41. **Gabon** (2) *GPS_Africa(2)*
42. **Germany** (26) *CALLISTO(2), SID(24)*
43. **Ghana** (1) *GPS_Africa(1)*
44. **Greece** (6) *AWESOME(1), SID(5)*
45. **Guyana** (2) *SCINDA(1), SID(1)*
46. **India** (25) *AWESOME(2), CALLISTO(4), CSSTE(1), MAGDAS(1), SEVAN(1), SID(16)*
47. **Indonesia** (10) *MAGDAS(8), SID(2)*
48. **Ireland** (14) *AWESOME(1), CALLISTO(4), SID(9)*
49. **Israel** (4) *AWESOME(1), ULF_ELF_VLF(3)*

50. **Italy** (41) *CALLISTO(2), MAGDAS(1), SID(38)*
51. **Japan** (13) *CHAIN(1), GMDN(1), MAGDAS(7), OMTIs(4)*
52. **Jordan** (1) *CSSTE(1)*
53. **Kazakhstan** (1) *CALLISTO(1)*
54. **Kenya** (8) *CALLISTO(1), GPS_Africa(1), MAGDAS(1), SCINDA(2), SID(3)*
55. **Korea** (2) *SID(2)*
56. **Kuwait** (1) *GMDN(1)*
57. **Lebanon** (11) *SID(11)*
58. **Libya** (3) *AWESOME(2), SID(1)*
59. **Madagascar** (1) *MAG_Africa(1)*
60. **Malaysia** (23) *AWESOME(1), CALLISTO(3), MAGDAS(2), OMTIs(1), SID(16)*
61. **Mali** (4) *GPS_Africa(2), MAG_Africa(2)*
62. **Mauritius** (3) *CALLISTO(3)*
63. **Mexico** (17) *CALLISTO(1), CSSTE(1), SAVNET(1), SID(14)*
64. **Micronesia** (1) *MAGDAS(1)*
65. **Mongolia** (13) *CALLISTO(2), MAGDAS(1), SID(10)*
66. **Morocco** (22) *AWESOME(1), CSSTE(1), GPS_Africa(19), RENOIR(1)*
67. **Mozambique** (4) *GPS_Africa(1), MAGDAS(1), SID(2)*
68. **Namibia** (4) *AMBER(1), GPS_Africa(1), MAG_Africa(1), SID(1)*
69. **Netherlands** (2) *SID(2)*
70. **New Zealand** (4) *SID(4)*
71. **Niger** (1) *GPS_Africa(1)*
72. **Nigeria** (47) *AMBER(1), CSSTE(1), MAGDAS(3), SCINDA(4), SID(38)*
73. **Norway** (1) *OMTIs(1)*
74. **Pakistan** (3) *SID(3)*
75. **Peru** (8) *CHAIN(1), CIDR(1), MAGDAS(2), SAVNET(3), SCINDA(1)*
76. **Philippines** (8) *MAGDAS(6), SCINDA(1), SID(1)*
77. **Poland** (1) *AWESOME(1)*
78. **Portugal** (3) *SID(3)*
79. **Rep of Congo** (1) *SID(1)*
80. **Republic of the Marshal Islands** (1) *SCINDA(1)*
81. **Romania** (3) *SID(3)*
82. **Russia** (13) *CALLISTO(1), MAGDAS(9), OMTIs(2), SID(1)*
83. **Sao Tome and Principe** (2) *GPS_Africa(1), SCINDA(1)*
84. **Scotland** (1) *SID(1)*
85. **Senegal** (3) *GPS_Africa(1), MAG_Africa(1), SID(1)*
86. **Serbia** (2) *AWESOME(1), SID(1)*
87. **Slovakia** (4) *CALLISTO(1), SEVAN(1), SID(2)*
88. **Slovenia** (1) *SID(1)*
89. **South Africa** (26) *GPS_Africa(7), MAGDAS(2), MAG_Africa(2), SCINDA(2), SID(13)*
90. **South Korea** (2) *CALLISTO(2)*
91. **Spain** (4) *CALLISTO(2), MAG_Africa(1), SID(1)*
92. **Sri Lanka** (2) *CALLISTO(1), SID(1)*
93. **Sudan** (1) *MAGDAS(1)*
94. **Sweden** (3) *SID(3)*
95. **Switzerland** (8) *CALLISTO(5), SID(3)*
96. **Taiwan** (4) *MAGDAS(1), SID(3)*
97. **Tanzania** (2) *MAGDAS(1), SCINDA(1)*
98. **Thailand** (6) *OMTIs(1), SID(5)*
99. **Tunisia** (8) *AWESOME(1), SID(9)*
100. **Turkey** (3) *AWESOME(1), SID(2)*
101. **UAE** (1) *AWESOME(1)*
102. **UK** (20) *CALLISTO(1), MAG_Africa(1), SCINDA(2), SID(16)*
103. **URUGUAY** (4) *SID(4)*

104. **USA** (270) *AWESOME(2), CALLISTO(3), CIDR(9), MAGDAS(2), SCINDA(2), SID(252)*
105. **US Virgin Islands** (2) *SID(2)*
106. **Uganda** (5) *SCINDA(1), SID(4)*
107. **Ukraine** (1) *CALLISTO(1)*
108. **Uruguay** (3) *SID(3)*
109. **Uzbekistan** (3) *AWESOME(1), SID(2)*
110. **Venezuela** (2) *SID(2)*
112. **Viet Nam** (3) *AWESOME(1), MAGDAS(1), SID(1)*
112. **Zambia** (3) *MAGDAS(1), SID(2)*

## LEGEND

**AMBER** African Meridian B-field Education and Research
**AWESOME** Atmospheric Weather Education System for Observation and Modelling of Effects
**CALLISTO** Compound Astronomical Low-cost Low-frequency Instrument for Spectroscopy and Transportable Observatory
**CHAIN** Continuous H-alpha Imaging Network
**CIDR** Coherent Ionospheric Doppler Radar
**CSSTE** Centres for Space Science and Technology Education
**GMDN** Global Muon Detector Network
**GPS Africa** African Dual Frequency GPS Network
**MAGDAS** Magnetic Data Acquisition System
**MAG Africa** Magnetometers in Africa
**OMTIs** Optical Mesosphere Thermosphere Imager
**RENOIR** Remote Equatorial Nighttime Observatory for Ionospheric Regions
**SAVNET** South America Very Low frequency Network
**SCINDA** Scintillation Network Decision Aid
**SEVAN** Space Environment Viewing and Analysis Network
**SID** Sudden Ionospheric Disturbance Monitor
**ULF_ELF_VLF** ULF/ELF/VLF network

# Current distribution of the International Space Weather Initiative instrument arrays as operational data sources

### Distribution of instruments by country

### Centres for Space Science and Technology Education (CSSTE)
6 centres, hosted in 6 countries
Brazil (1), India (1), Jordan (1), Mexico (1), Morocco (1), Nigeria (1),

### African Meridian B-field Education and Research (AMBER)
5 instruments, hosted in 5 countries
Algeria (1), Cameroon (1), Ethiopia (1), Namibia (1), Nigeria (1),

### Atmospheric Weather Education System for Observation and Modelling of Effects (AWESOME)
24 instruments, hosted in 21 countries
Algeria (1), Antarctica (1), Azerbaijan (1), Egypt (1), Ethiopia (1), Fiji (1), Greece (1), India (2), Ireland (1), Israel (1), Libya (2), Malaysia (1), Morocco (1), Poland (1), Serbia (1), Tunisia (1), Turkey (1), UAE (1), USA (2), Uzbekistan (1), Viet Nam (1),

**Compound Astronomical Low-cost Low-frequency Instrument for Spectroscopy and Transportable Observatory (CALLISTO)**

51 instruments, hosted in 27 countries
Australia (2), Austria (1), Belgium (1), Brazil (2), Costa Rica (1), Czech Republic (1), Egypt (1), Finland (2), Germany (2), India (4), Ireland (4), Italy (2), Kazakhstan (1), Kenya (1), Malaysia (3), Mauritius (3), Mexico (1), Mongolia (2), Russia (1), Slovakia (1), South Korea (2), Spain (2), Sri Lanka (1), Switzerland (5), UK (1), USA (3), Ukraine (1),

**Continuous H-alpha Imaging Network (CHAIN)**

3 instruments, hosted in 3 countries
Algeria (1), Japan (1), Peru (1),

**Coherent Ionospheric Doppler Receivers (CIDR)**

11 instruments, hosted in 3 countries
Egypt (1), Peru (1), USA (9),

**Global Muon Detector Network (GMDN)**

4 instruments, hosted in 4 countries
Australia (1), Brazil (1), Japan (1), Kuwait (1),

**African Dual Frequency GPS Network (GPS_Africa)**

43 instruments, hosted in 16 countries
Algeria (1), Benin (1), Botswana (1), Burkina Faso (2), Cape Verde (1), Gabon (2), Ghana (1), Kenya (1), Mali (2), Morocco (19), Mozambique (1), Namibia (1), Niger (1), Sao Tome (1), Senegal (1), South Africa (7),

**Magnetic Data Acquisition System (MAGDAS)**

71 instruments, hosted in 27 countries
Australia (11), Brazil (2), Canada (1), Cote d'Ivoire (1), Ecuador (1), Egypt (2), Ethiopia (1), India (1), Indonesia (8), Italy (1), Japan (7), Kenya (1), Malaysia (2), Micronesia (1), Mongolia (1), Mozambique (1), Nigeria (3), Peru (2), Philippines (6), Russia (9), South Africa (2), Sudan (1), Taiwan (1), Tanzania (1), USA (2), Viet Nam (1), Zambia (1),

**Magnetometers in Africa (MAG_Africa)**

14 instruments, hosted in 11 countries
Algeria (1), Central African Rep. (1), Cote d'Ivoire (2), Ethiopia (1), Madagascar (1), Mali (2), Namibia (1), Senegal (1), South Africa (2), Spain (1), UK (1),

**Optical Mesosphere Thermosphere Imager (OMTIs)**

12 instruments, hosted in 7 countries
Australia (1), Canada (2), Japan (4), Malaysia (1), Norway (1), Russia (2), Thailand (1),

**Remote Equatorial Nighttime Observatory for Ionospheric Regions (RENOIR)**

3 instruments, hosted in 2 countries
Brazil (2), Morocco (1),

**South America Very Low frequency Network (SAVNET)**

11 instruments, hosted in 4 countries
Argentina (1), Brazil (6), Mexico (1), Peru (3),

**Scintillation Network Decision Aid (SCINDA)**

    34 instruments, hosted in 22 countries
    Brazil (3), Cameroon (1), Cape Verde (1), Chile (1), Colombia (1), Congo (3),
    Cote d'Ivoire (1), Djibouti (1), Egypt (1), Ethiopia (2), Guyana (1), Kenya (2),
    Nigeria (4), Peru (1), Philippines (1), Rep. of Marshal Islands (1),
    Sao Tome and Principe (1), South Africa (2), Tanzania (1), UK (2), USA (2),
    Uganda (1),

**Space Environment Viewing and Analysis Network (SEVAN)**

    7 instruments, hosted in 5 countries
    Armenia (3), Bulgaria (1), Croatia (1), India (1), Slovakia (1),

**Sudden Ionospheric Disturbance Monitor (SID)**

    657 instruments, hosted in 75 countries
    Algeria (2), Antarctica (1), Australia (3), Austria (2), Azerbaijan (2),
    Bangladesh (1), Bosnia-Herzegovina (1), Brazil (7), British Virgin Islands (1),
    Bulgaria (1), Burkina Faso (1), CHINA (3), Canada (17), China (26),
    Colombia (7), Congo (4), Croatia (5), Cypress (1), Czech Republic (1),
    D Rep of Congo (2), Denmark (2), Egypt (3), England (1), Ethiopia (14),
    France (4), Germany (24), Greece (5), Guyana (1), India (16), Indonesia (2),
    Ireland (9), Italy (38), Kenya (3), Korea (2), Lebanon (11), Libya (1),
    Malaysia (16), Mexico (14), Mongolia (10), Mozambique (2), Namibia (1),
    Netherlands (2), New Zealand (4), Nigeria (38), Pakistan (3), Philippines (1),
    Portugal (3), Rep of Congo (2), Romania (3), Russia (1), Scotland (1),
    Senegal (1), Serbia (1), Slovakia (2), Slovenia (1), South Africa (13),
    Spain (1), Sri Lanka (1), Sweden (3), Switzerland (3), Taiwan (3),
    Thailand (5), Tunisia (7), Tunisia (2), Turkey (2), UK (16), URUGUAY (4),
    USA (252), US Virgin Islands (2), Uganda (4), Uruguay (3), Uzbekistan (2),
    Venezuela (2), Viet Nam (1), Zambia (2),

**ULF/ELF/VLF network (ULF_ELF_VLF)**

    3 instruments, hosted in 1 countries
    Israel (3),

## Published scientific papers based on results of the International Space Weather Initiative in the period of time from 2009 - 2013

Table 4

**2013**

1.    Raulin, J.-P., G.H.Trottet, M.Kretzschmar, E.L.Macotela, A.A.Pacini, F.C.P.Bertoni, I.Dammasch Response of the low ionosphere to X-ray and Lyman-alpha solar flare emissions (2013) Journal of Geophysical Research, Article in Press (ADS) doi: 10.1029/2012JA017916

2.    De La Luz, V., Raulin, J.-P., Lara, A. The Chromospheric Solar Millimeter-wave Cavity originates in the temperature minimum region (2013) Astrophysical Journal, 762 (2), art. no. 84, Article

3.    Correia, E., Raulin, J.-P., Kaufmann, P., Bertoni, F., Quevedo, M.T. Inter-hemispheric analysis of daytime low ionosphere behavior from 2007 to 2011 (2013) Journal of Atmospheric and Solar-Terrestrial Physics, 92, pp. 51-58. Article

**2012**

## 2011

Science Initiative: 2010 Status Report on the International Space Weather Initiative (2011) Sun and Geosphere, vol.6, no.1, p.7-16. (ADS)

2007 campaign (2010) Journal of Geophysical Research, Volume 115, Issue 1, CiteID A00E47 (ADS) doi: 10.1029/2009JA015026

Highlights of Astronomy, Volume 15, p. 480-483 (ADS) doi: 10.1017/S1743921310010331

**2009**

Advances in Space Research, Volume 43, Issue 4, p. 717-720. (ADS) doi: 10.1016/j.asr.2008.10.008

282. Chilingarian, A., Mirzoyan, R., Zazyan, M. Cosmic Ray research in Armenia (2009) Advances in Space Research, Volume 44, Issue 10, p. 1183-1193. (ADS) doi: 10.1016/j.asr.2008.11.029

283. Chilingarian, A., Hovsepyan, G., Arakelyan, K., Chilingaryan, S., Danielyan, V., Avakyan, K., Yeghikyan, A., Reymers, A., Tserunyan, S. Space environmental viewing and analysis network (SEVAN) (2009) Earth, Moon and Planets, 104 (1-4), pp. 195-210. Cited 5 times. Conference Paper doi: 10.1007/s11038-008-9288-1

284. Carrano, C.S., Yizengaw, E., Doherty, P.H., Bridgwood, C.T., Adeniyi, J.O., Amaeshi, L.L., Pedersen, T.R., Groves, K.M., Roddy, P.A., Caton, R.G. Ground- and Space-Based Observations of Ionospheric Irregularities over Nigeria during Solar Minimum (2009) American Geophysical Union, Fall Meeting 2009, abstract #SA31B-1424 (ADS)

285. Chilingaryan, S., Chilingarian, A., Danielyan, V., Eppler, W. Advanced data acquisition system for SEVAN (2009) Advances in Space Research, 43 (4), pp. 717-720. Cited 2 times. Article

286. Chilingarian, A., Bostanjyan, N. Cosmic ray intensity increases detected by Aragats Space Environmental Center monitors during the 23rd solar activity cycle in correlation with geomagnetic storms (2009) Journal of Geophysical Research, Volume 114, Issue A9, CiteID A09107 (ADS) doi: 10.1029/2009JA014346

287. Chilingarian, Ashot Statistical study of the detection of solar protons of highest energies at 20 January 2005 (2009) Advances in Space Research, Volume 43, Issue 4, p. 702-707. (ADS) doi: 10.1016/j.asr.2008.10.005

288. Bong, Su-Chan, Kim, Yeon-Han, Roh, Heeseon, Cho, Kyung-Suk, Park, Young-Deuk, Choi, Seonghwan, Baek, Ji-Hye, Monstein, Christian, Benz, Arnold O., Moon, Yong-Jae, Kim, Sungsoo S. Constructino of an E-Callisto Station in Korea (2009) Journal of the Korean Astronomical Society, vol. 42, no. 1, pp. 1-7 (ADS)

289. Bisi, Mario, Jackson, B.V., Hick, P.P.L., Clover, J.M., Tokumaru, M., Fujiki, K., Fallows, R.A., Breen, A.R. Three-Dimensional Reconstructions of the Solar Wind: During Solar Minimum Conditions (2009) Bulletin of the American Astronomical Society, Vol. 41, p.866 (ADS)

290. Barton, C.E., C.Amory-Mazaudier, B.Barry, V.Chukwuma, R.L.Cottrell, U.Kalim, A.Mebrathu, M.Petitdidier, B.Rabiu, C.Reeves Electronic Geophysical Year: Start of the Art and Results: eGY-Africa: Addressing the digital divide for science in Africa (2009) Russian Journal of Earth Sciences, Vol. 11, ES003 (ADS)

291. Benz, A.O., Monstein, C., Meyer, H., Manoharan, P.K., Ramesh, R., Altyntsev, A., Lara, A., Paez, J., Cho, K.-S. A World-Wide Net of Solar Radio Spectrometers: e-CALLISTO (2009) Earth, Moon, and Planets, Volume 104, Issue 1-4, pp. 277-285 (ADS) doi: 10.1007/s11038-008-9267-6

292. Benz, Arnold O., Monstein, Christian, Beverland, Michael, Meyer, Hansueli, Stuber, Bruno High Spectral Resolution Observation of

## 5. Continuation of the Basic Space Science Initiative Beyond 2012

The Basic Space Science Initiative will continue providing support to operators of planetariums, astronomical telescopes, and ISWI instruments.

### Astronomical telescopes

The International Scientific Optical Network (ISON) is an open international non-government project mainly aimed at being a free source of information on space objects for scientific analysis and other applications. It was initiated in the framework of the programme of the GEO region investigations started by the Keldysh Institute of Applied Mathematics (KIAM) of the Russian Academy of Sciences in 2001 and in order to support space debris radar experiments with additional tracking data used for determination of orbital parameters precise enough to properly point narrow radar beams of selected objects.

ISON is one of the largest observation systems and it is one of two existing systems in the world, capable to observe the sky globally from both — Eastern and Western — hemispheres. At present, there are more than 30 telescopes operating in 20 observatories in eight nations: Bolivia, Georgia, Italy, Moldova, Russia, Tajikistan, Ukraine, Uzbekistan. All these telescopes participate in a coordinated observation programme under the ISON project.

ISON telescopes are grouped in three subsets dedicated to tracking of different classes of the space objects – bright GEO-objects, faint fragments at GEO region, bright objects at highly elliptical (HEO), and low orbits (LEO). ISON activities are arranged with four supporting groups such as (i) electric and software engineering, (ii) optical and mount engineering, (iii) observation planning and data processing, and (iv) network development. The obtained data are collected and stored at the KIAM Centre for Processing and Analysis of Information on Space Debris (CCPAISD), Russian Academy of Sciences.

The goal of the ISON observations of faint space debris fragments at high orbits was formulated from the beginning of the establishment of ISON in 2004. First experiments arranged with a 64 cm telescope AT-64 in Nauchny, Crimea, in October 2004 were devoted to adjusting of a method of the fragments discovering and checking of the Pulkovo theory on orbital evolution of the GEO object explosion fragments. These successful attempts, discovering seven not catalogued fragments and obtaining 1240 measurements in 18 tracks, initiated also the cooperation with a team at the Astronomical Institute of the University of Bern (AIUB), Switzerland. The regularly coordinated AIUB-ISON observing campaigns were carried out during 2005 and the ISON subsystem for the tracking of the faint fragments at GEO region started operations in 2006. ISON news are regularly published in a dedicated website: www.lfvn.astronomer.ru.

### Planetariums

The recently published book titled Max Goes to the Moon has been donated by author Jeffrey Bennett (www.jeffreybennett.com) and

provided to public educational institutions by the Office for Outer Space Affairs of the United Nations (www.unoosa.org). Max Goes to the Moon was the first children's book selected to be read aloud in space, by Astronaut Alvin Drew aboard the final mission of the Space Shuttle Discovery. To see a video of the reading, go to www.bigkidscience.com/max_in_space. Planetariums in your region can obtain materials for a Max Goes to the Moon planetarium show at no cost. A variety of other free educational resources related to this and other books in the Max series are available at www.BigKidScience.com. Jeffrey Bennett also distributes e-mails with educational information about space science. Educational institutions that are using the book are invited to provide brief information on ways and means to incorporate the book into their educational activities to the Office for Outer Space Affairs of the United Nations.

**Space weather instruments**

In 2012, the Scientific and Technical Subcommittee of COPUOS, at its forty-ninth session, agreed that an agenda item entitled "Space Weather" should be introduced as a regular item on the agenda of the Subcommittee, in order to allow member States of COPUOS and international organizations having permanent observer status with COPUOS to exchange views on national, regional, and international activities related to space weather research with a view to promoting greater international cooperation in that area. The Subcommittee noted that it could, through that item, serve as an important advocate for efforts to close existing gaps in the space weather research field (A/AC.105/1001, para. 226).

Space weather is important to society, which increasingly relies on technology for education, business, transportation and communication. Space storms can disrupt GPS/GNSS reception and long distance radio transmission. Modern oil and gas drilling frequently involve directional drilling to tap oil and gas reservoirs deep in the Earth, thus depending on accurate positioning using GPS systems. Energetic particles at the magnetic poles can force re-routing of polar airline flights resulting in delays and increased fuel consumption. Ground induced currents generated by magnetic storms can cause extended power blackouts and increased corrosion in critical energy pipelines.

In addition, space weather likely affects Earth's climate. The amount of energy entering the troposphere and stratosphere from all space weather phenomena is trivial compared to the solar insolation in the visible and infrared portions of the solar electromagnetic spectrum. However, there does seem to be some linkage between the 11-year sunspot cycle and Earth's climate. For example, the Maunder minimum, a 70 year period almost devoid of sunspots, correlates with a cooling of Earth's climate.

Space weather is inherently an international enterprise. Solar and magnetic storms can affect large regions of the Earth simultaneously, and equatorial ionospheric disturbances occur routinely around the globe. It is therefore appropriate for the United Nations to promote improvement in space weather modelling and forecasting for the benefit of all nations.

There has been significant scientific progress over the past decade in developing physics-based space weather models, and large-scale coupled (real-time) space plasma simulations. However, these models are limited by still being data starved in various spatial space weather domains. Thus, there is a crucial need for guaranteed continuous space weather data streams.

In 2007, the International Heliophysical Year (IHY) and International Space Weather Initiative (ISWI) have made significant progress in the installation of new instrumentation for the understanding of space weather impacts on Earth's upper atmosphere, generating new data streams useful for space weather in regions unobserved before. With the support of the United Nations Office of Outer Space Affairs (OOSA), the ISWI has facilitated the operation of nearly 1000 instruments operating in 100 of the UN's 194 Member States. The data from these instrument arrays is a unique resource for the study of space weather influences on Earth's atmosphere. The IHY and ISWI schools have trained several hundred graduate students and young scientists, many of whom are maturing as fine scientists as evidenced by their publications.

**ISWI Scientific Objectives and Programme**

At the 2011 UN/Nigeria Workshop on ISWI (A/AC.105/1018), the desire for a programme to enhance the science from ISWI was announced. The announcement also appeared at that time in the ISWI Newsletter (http://iswi-secretariat.org/). This was followed up by a presentation and subsequent discussion of an ISWI science plan at the 2012 UN/Ecuador Workshop on ISWI (A/AC.105/1030), resulting in the following observations and recommendations.

The overall ISWI objective is to develop the scientific insight necessary to understand the science, and reconstruct and forecast near-Earth space weather. Steps in this process include: (i) Expanding and continuing the deployment of the existing ISWI instrument arrays and add new arrays as appropriate. (ii) For the data being obtained by the arrays, expanding the data analysis effort for array data and using existing data bases. (iii) Coordinating the data products to provide inputs for physical modelling, input instrument array data into physical models of heliospheric processes, and developing data products that reconstruct past conditions in order to facilitate assessment of problems attributed to space weather effects. (iv) Coordinating the data products to permit predictive relationships to be developed, to develop data products yielding predictive relationships that enable the forecasting of Space Weather, and to develop data products that can easily be assimilated into real-time or near real-time predictive models.

Fundamental aspects of ISWI include education and training, and public outreach activities. The concept is to encourage and support space science courses, workshops and curricula in university and graduate schools that provide instrument support. There has been much success in these areas, but there is a strong need to continue the education and training, develop public outreach materials that are unique to ISWI, and coordinate their distribution. It is important to provide information on ISWI instruments and results to the media, especially to local media.

The overall goal of an ISWI science programme should be to gain a more complete understanding of the universal processes that govern the Sun, Earth, planets, and heliosphere. This must involve scientists from a variety of disciplines, such as Solar Physics, Planetary Magnetospheres, Heliosphere and Cosmic Rays, Planetary Ionospheres, Thermospheres and Mesospheres, and Climate Studies. ISWI science projects should focus on the fundamental underlying physics of each phenomenon. They should facilitate discussions between different disciplines by focusing on relationships between these phenomena and commonalities in the physical processes. This allows researchers to plan and participate in cross-disciplinary studies, culminating in a greater understanding of fundamental universal processes.

What are the science benefits? This question is informed by recognizing what is unique about the ISWI data sets: (i) By observing in new geographical regions, a more global picture of Earth's response to solar wind inputs can be obtained; (ii) 24/7 solar observing is obtained in the radio and H-alpha wavelength regimes; (iii) Arrays provide global, 3-D information that can be used in tomographic reconstructions; (iv) Over the long term, these arrays will provide real-time data valuable for forecasting and now casting; and (v) Modelling improvements will allow better exploitation of existing data sets.

*Enhancing the Science from ISWI*

The primary data sets for ISWI science should come from the ISWI (IHY Legacy) Instrument Projects/Arrays. There are currently 16 operational instrument arrays hosted by ~100 nations. Current scientific activities sponsored by ISWI will continue. Specific recommendations are:

1. The ISWI will continue the deployment of instruments both for existing instrument arrays and for new instrument arrays on a perpetual basis.

2. The ISWI will undertake a process to use ISWI instrument data sets to determine data utility, to develop connections with virtual observatories to make data more readily available, and to facilitate collaborative modelling of regions of interest (e.g., the equatorial ionosphere) in collaboration with modelling centres of the ESA JAXA, NASA and others.

3. Data from ISWI instrument arrays will be combined with space-based data to advance space weather science leading to quality papers in international journals.

4. Space Science Schools are an integral part of ISWI, providing training for young and new researchers in instrument operation and the science of heliophysics. ISWI Space Science Schools will continue, and partnerships with organizations such as SCOSTEP and the International Astronomical Union will be established or strengthened to assure that these capacity building activities are accomplished efficiently and for the benefit of all member States.

Another new aspect of ISWI might include coordinated programmes or campaigns modelled on the Coordinated Investigation Programmes (CIPs) established under the International Heliophysical Year (IHY). These could use the same principles as the IHY scientific activities, but need not be as formal as the CIPs were. General CIP science topics included cosmic rays, solar filaments, CME onset and propagation, incoherent scatter radar data and comparative aeronomy.

The main communication interface for ISWI science should be an internet site set up either directly through or via a link to the main ISWI website at www.iswi-secretariat.org/. The science program can aid in establishing space weather modelling centres around the world, e.g., in collaboration with NASA's Coordinated Community Modeling Center (CCMC) at http://ccmc.gsfc.nasa.gov/. It would provide an interface for ISWI science results with the ISWI Space Science Schools and Public Outreach projects. Specific details on developments of the ISWI science programme will be made available through the ISWI Newsletter.

One way of developing and promoting scientific results from ISWI could be to have one or more ISWI science workshops, the first possibly as early as next year, 2013. The workshop would focus on scientific results. This or a follow on workshop could be like a NASA Coordinated Data Analysis Workshop (CDAW) in which the data interfaces and analysis tools are prepared in advance, and the results are presented at one or more later meetings and then published together in one volume.

*Developing Data Analysis Links between ISWI Projects and/or with Outside Projects*

In the following, some recent examples of cooperative science projects already underway or planned that involve ISWI array data. These include some examples of combining data from several ISWI arrays and with other "outside" data sources.

The Continuous H-alpha Imaging Network (CHAIN) project, with PIs K. Shibata and S. Ueno of Japan, has the goal of forming an worldwide H-alpha observing network for understanding and predicting space weather by accurately observing erupting phenomena on the solar surface that are initial boundary conditions of all eruptive processes. Its scientific goals are determination of the 3-D velocity field of eruptive phenomena on the solar surface, detection of shock waves (Moreton waves) generated by solar explosive phenomena, and estimation of solar UV radiation and comparison with ionospheric variation. The CHAIN group is seeking partnerships for using these data in international cooperative studies. For example, this year they plan to work with the Pakistan Space and Upper Atmosphere Research Commission (SUPARCO) to study F2 ionospheric density variations during solar minimum and maximum conditions, ionospheric variability at low and mid latitudes for solar cycles 22and 23, and solar cycle effects on coupling of neutral and ionized species at F2 altitude.

The Compact Astronomical Low-cost Low-frequency Instrument for Spectroscopy in Transportable Observatories (CALLISTO), PIs A. Benz and C. Monstein (ETHZ, Switzerland), is a radio spectrometer using a heterodyne receiver build by the ETH Zürich Radio and Plasma Physics Group. CALLISTO is able to continuously cover the solar radio spectrum from 45 to 870 MHz, using modern, commercially available broadband cable-TV tuners having frequency resolution of 62.5 KHz. CALLISTO has now deployed ~56 instruments in more than 30 locations with users from more than 74 countries. It produces science quality data and detects even tiny eruptions from the Sun. The spectral data on solar radio bursts is being used by the solar community but needs wider distribution. CALLISTO data is now being utilized for an Indo-US project on solar eruptive events. With this large network CALLISTO provides 24/7 coverage of radio bursts. Future plans include identifying sets of similar instruments to provide more continuous coverage over the whole frequency range.

The African Meridian B-Field Education and Research (AMBER), PI's E. Yizengaw (Boston College, USA) and M. Moldwin (University of Michigan, USA), is a magnetometer array comprised of four magnetometers stationed in Ethiopia, Algeria, Cameroon, and in Namibia. AMBER's data are being combined with other related arrays to provide important new observations for these objectives: (i) To monitor the electrodynamics governing the motion of plasma in the low/mid-latitude as function of LT, season, and magnetic activity; (ii) To understand ULF pulsation strength into low/mid-latitudes and its connection with EEJ and EA; (iii) To support studies about the effects of Pc5 ULF waves on the MeV electron population in the inner parts of the Van Allen belts. Recently, in collaboration with the SAMBA (E. Zesta, AFRL, USA) project, a fifth magnetometer was deployed in Nigeria. Other non-ISWI networks that are being used with AMBER are LISN
(C. Valladares, Boston College, USA) in South America and MEASURE (M. Moldwin) in North America. AMBER is working to coordinate with other ground-based magnetometer arrays to provide a worldwide network to understand the electrodynamics that governs equatorial ionosphere motions. Data from the AMBER and related magnetometer arrays will be made accessible to space weather forecasters and the space science community at large.

There are several international programmes for which the use of ISWI data is planned. The International Study of Earth-Effecting Solar Transients (ISEST) is coordinated by Jie Zhang of George Mason University, USA, and is being implemented in 2012-2013 under the framework of Task Group 3 of SCOSTEP/CAWSES II. Its goals are: (i) To organize three international workshops; (ii) To promote an international effort to create a comprehensive database of Earth-effecting solar and heliospheric transient events in solar cycle 23 and 24, to develop advanced theoretical models of heliospheric transients, and to develop prediction tools for heliospheric transients. Another program is an ISWI Study of Radio Transients coordinated by N. Gopalswamy of the ISWI Secretariat. Two other international programmes that D. Webb coordinates are the IAU Working Group on International Collaboration on Space Weather and the STEREO mission Space Weather Group. Both of these could utilize ISWI data.

Table 4. lists published scientific papers that used ISWI array data using the Astrophysics Data System (ADS).

*Conclusions*

Space weather research is beneficial to the technological society. The ISWI makes an important and unique contribution by developing new data sources necessary for improved space weather understanding and prediction. The initial emphasis in ISWI has been on studies with analyses of data from individual ISWI instrument arrays. It is now time to expand these studies and it is recommended that, where possible and scientifically justified, these projects include cooperative studies with data sets from other ISWI arrays and other "outside" data sets to best address and expand the science results. Above a few examples of such expanded projects utilizing data from the ISWI arrays are given. It is recommended that these projects continue to interface with the ISWI Space Science Schools, Public Outreach projects and other educational science programmes. The ISWI community at large also would like to receive feedback from the PI groups of the instrument arrays and projects on how best to proceed on enhancing ISWI science. It is expected that any new data plan or procedures will be part of the proposed new permanent agenda item "Space Weather" for the Scientific and Technical Subcommittee of COPUOS.

*Further Reading*

D.W. Smith and H.J. Haubold (Eds.), Planetarium: A Challenge for Educators, United Nations, New York 1992, www.unoosa.org/pdf/publications/planetariumE.pdf

W. Wamsteker, R. Albrecht, and H.J. Haubold (Eds.), Developing Basic Space Science World-Wide: A Decade of UN/ESA Workshops, Kluwer Academic Publishers, Dordrecht-Boston-London 2004.

B.J. Thompson, N. Gopalswamy, J.M. Davila, and H.J. Haubold (Eds.), Putting the "I" in IHY: The United Nations Report for the International Heliophysical Year 2007, Springer, Wien-New York 2009.

The Working Group on Space Sciences in Africa (www.saao.ac.za/~wgssa/) is an international, non-governmental organisation founded by African delegates at the 6th United Nations/European Space Agency Workshop held in Bonn on 9-13 September 1996. The scientific scope of the Working Group's activities is defined to encompass: astronomy and astrophysics, solar-terrestrial interaction and its influence on terrestrial climate, planetary and atmospheric studies, and the origin of life and exobiology. The Working Group seeks to promote the development of the space sciences in Africa by initiating and coordinating various capacity-building programmes throughout the region. These programmes fall into a broad spectrum ranging from the promotion of basic scientific literacy in the space sciences to the support of international research projects. The Working Group also promotes international cooperation among African space scientists and acts as a forum for the exchange of ideas and information through its publications, outreach programmes, workshops, and scientific meetings. Since the above workshop, WGSSA is publishing an annual volume on new developments in astronomy in Africa, including achievements of the basic space science initiative (www.saao.ac.za/~wgssa/archive.php).

Acknowledgements: The authors are grateful to all who have joined the long process of implementation of the UNBSSI since 1988, especially those who took part in the workshops. Special thanks to ESA, NASA, and JAXA for all support provided. Committees established for IHY 2007 and ISWI have been very instrumental to the success of the initiative. Host countries of the workshops and their respective Governments and expert groups provided enlightening support and guidance for the cultural and scientific atmosphere to the whole initiative.

Disclaimer: The views expressed in this paper are solely those of the authors and do not necessarily reflect the views of any company or organization.